\pdfoutput=1
\documentclass[apj]{emulateapj}
\usepackage{natbib}
\usepackage{multirow}
\usepackage[pdftex,                
bookmarks=true,                    
bookmarksnumbered=true,            
colorlinks=true,            
citecolor=blue,         
pagecolor=blue,         
linkcolor=blue,     
menucolor=blue,         
urlcolor=cyan,       
linkbordercolor={0 0 1},           
pdfborder= {0 0 1},
frenchlinks=True]{hyperref}

\providecommand{\eprint}[1]{\href{http://arxiv.org/abs/#1}{#1}}
\providecommand{\adsurl}[1]{\href{#1}{ADS}}
\providecommand{\emailh}[1]{\protect{\href{mailto:#1}{#1}}}
\def\aap{{A\&A}}		
\def\apj{{ApJ}}			
\def\apjl{{ApJ}}		
\def\apjs{{ApJS}}		
\def\pasp{{PASP}}		

\def\mnras{{MNRAS}}
\def\nat{{Nature}}

\shortauthors{Crossfield et al.}
\shorttitle{New $\upsilon$~And~b Phase Curve}
\begin{document}

\newcommand{\tone}{\ensuremath{T_{P1}}}
\newcommand{\ttwo}{\ensuremath{T_{P2}}}
\newcommand{\slab}[1]{\label{sec:#1}}
\newcommand{\sref}[1]{Sec.~\ref{sec:#1}}
\newcommand{\flab}[1]{\label{fig:#1}}
\newcommand{\fref}[1]{Figure~\ref{fig:#1}}
\newcommand{\elab}[1]{\label{eq:#1}}
\newcommand{\eref}[1]{Eq.~(\ref{eq:#1})}
\newcommand{\erefs}[3]{Eqs.~(\ref{eq:#1})#2(\ref{eq:#3})}
\newcommand{\tlab}[1]{\label{tab:#1}}
\newcommand{\tref}[1]{Table~\ref{tab:#1}}
\newcommand{\ua}{\ensuremath{\upsilon}~And}
\newcommand{\uab}{\ensuremath{\upsilon}~And~b}
\newcommand{\uand}{\ensuremath{\upsilon}~Andromedae}
\newcommand{\uandb}{\ensuremath{\upsilon}~Andromedae~b}
\newcommand{\hdtwo}{HD~209458}
\newcommand{\hdtwob}{HD~209458b}
\newcommand{\hdone}{HD~189733}
\newcommand{\hdoneb}{HD~189733b}
\newcommand{\msun}{\ensuremath{M_\odot}}
\newcommand{\rsun}{\ensuremath{R_\odot}}
\newcommand{\rjup}{\ensuremath{R_{J}}}
\newcommand{\mjup}{\ensuremath{M_{J}}}
\newcommand{\degg}{\ensuremath{^{\circ}}}
\newcommand{\figtwocol}[5]{
        \begin{figure*}[!bt]
        \begin{center}
        \includegraphics[#3]{#2}
        \end{center}
        \renewcommand{\baselinestretch}{1}
        \vspace*{-.3in}
        \caption[#4]{\footnotesize #5}
        \flab{#1}
        \end{figure*}}

\newcommand{\fig}[5]{
        \begin{figure}[!bt]
        \begin{center}
        \includegraphics[#3]{#2}
        \end{center}
        \renewcommand{\baselinestretch}{1}
        \vspace*{-.3in}
        \caption[#4]{#5}
        \flab{#1}
        \end{figure}}

      \title{A New 24\,\micron\ Phase Curve for $\upsilon$
        Andromedae~\lowercase{b}}

\slugcomment{Submitted to {\em ApJ} 2010 August 2.}

\author{Ian J. Crossfield\altaffilmark{1}, Brad
  M. S. Hansen\altaffilmark{1,2}, Joseph Harrington\altaffilmark{3},
  James Y-K. Cho\altaffilmark{4}, Drake Deming\altaffilmark{5}, Kristen
  Menou\altaffilmark{6}, Sara Seager\altaffilmark{7}}

\altaffiltext{1}{Department of Physics \& Astronomy, University of California Los Angeles, Los Angeles, CA 90095, USA; \emailh{ianc@astro.ucla.edu}}
\altaffiltext{2}{Institute of Geophysics \& Planetary Physics, University of California Los Angeles, Los Angeles, CA 90095, USA; \emailh{hansen@astro.ucla.edu}}
\altaffiltext{3}{Planetary Sciences Group, Department of Physics, University of Central Florida, Orlando, FL 32816-2385, USA; \emailh{jh@physics.ucf.edu}}
\altaffiltext{7}{School of Mathematical Sciences, Queen Mary, University of London, London E1 4NS, UK;  \emailh{J.Cho@qmul.ac.uk}}
\altaffiltext{4}{Planetary Systems Branch Code 693, NASA/Goddard Space Flight Center, Greenbelt, MD 20771, USA; \emailh{Leo.D.Deming@nasa.gov}}
\altaffiltext{6}{Department of Astronomy, Columbia University, New York, NY 10027, USA; \emailh{kristen@astro.columbia.edu}}
\altaffiltext{5}{Department of Earth, Atmospheric and Planetary Sciences, Department of Physics, Massachusetts Institute of Technology, Cambridge, MA 02139, USA; \emailh{seager@mit.edu}}

\begin{abstract}
  We report the detection of 24\,\micron\ variations from the
  planet-hosting \uand\ system consistent with the orbital periodicity
  of the system's innermost planet, \uab .  We find a peak-to-valley
  phase curve amplitude of 0.00130 times the mean system flux.  Using
  a simple model with two hemispheres of constant surface brightness
  and assuming a planetary radius of 1.3~\rjup\ gives a planetary
  temperature contrast of $\gtrsim 900$~K and an orbital inclination
  of $\gtrsim 28\degg$.  We further report the largest phase offset
  yet observed for an extrasolar planet: the flux maximum occurs
  $\sim80\degg$ before phase 0.5. Such a large phase offset is
  difficult to reconcile with most current atmospheric circulation
  models. We improve on earlier observations of this system in several
  important ways: (1) observations of a flux calibrator star
  demonstrate the MIPS detector is stable to $10^{-4}$ on long
  timescales, (2) we note that the background light varies
  systematically due to spacecraft operations, precluding use of this
  background as a flux calibrator (stellar flux measured above the
  background is not similarly affected), and (3) we calibrate for flux
  variability correlated with motion of the star on the MIPS detector.
  A reanalysis of our earlier observations of this system is
  consistent with our new result.
\end{abstract}


\section{Introduction}

The first thermal characterizations of highly irradiated extrasolar
planetary atmospheres were made by measuring the flux decrement that
occurs during secondary eclipse, when an extrasolar planet passes
behind its host star \citep{deming:2005, charbonneau:2005}.  This
decrement gives an estimate of the (hemisphere-averaged) temperature
of a planet's star-facing side at the time of eclipse and provides
insight into the energy budgets of these hot worlds.  Secondary
eclipse (or, occultation) observations have been widely
interpreted as indicating two types of planetary atmospheres, namely
planets with and without high-altitude temperature inversions
\citep{burrows:2008,fortney:2008}.  However, the number of free
parameters in current models is such that in many cases it is still
difficult to place strong constraints on a planet's atmospheric
structure with the few data points available for most systems
\citep{madhusudhan:2009}.

\cite{burrows:2008} and \cite{fortney:2008} suggested that sufficiently
high levels of irradiation prevent strong optically absorbing species
from condensing and ``raining out'' of the upper atmosphere of hot
Jupiters, thus directly linking high levels of incident stellar flux
to the presence of a temperature inversion.  However, subsequent
secondary eclipse measurements with Spitzer/IRAC have complicated the
picture and a straightforward connection between irradiation and
inversions now seems untenable.  For example, TrES-3b receives
substantially more flux that does \hdtwob , yet the latter has an
atmospheric inversion \citep{knutson:2008} while the former does not
\citep{fressin:2010}.  Thus, planetary classification will require
more subtlety than a simple critical-flux level model can provide.
\cite{knutson:2010} have recently suggested a correlation between
stellar activity and the absence of a temperature inversion: in this
hypothesis high-altitude absorbing species are photodissociated by the
ultraviolet flux from an active star.  It remains to be seen how this
theory addresses the issue of temporally variable stellar activity
\citep{shkolnik:2008}.

Phase curves provide complementary insights into planetary
atmospheres.  If a system's total (star plus planet) infrared flux varies
periodically and in phase with the planet's orbit, the variation can
be attributed to spatially nonuniform radiation emitted by the planet.
Such measurements have the potential to constrain the planet's
circulation and heat redistribution patterns.  If incident stellar
flux were instantaneously re-radiated by the planet, the hottest
region on the planet would be at the substellar point; such a phase
curve is said to have zero phase offset.  Nonzero phase offsets thus
imply heat transport around the planet; for example, by advection of
absorbed stellar energy by global winds \citep{showman:2009}, or by
heating induced by atmospheric gravity waves \citep{watkins:2010}.
Interpreting phase curves can be challenging because the brightest
atmospheric region will also depend on the opacity structure of the
atmosphere and the wavelength at which one observes the system.

The extrasolar planet \uandb\ (\uab ) is the first exoplanet for which
a phase curve was reported. \citeauthor{harrington:2006}
(\citeyear{harrington:2006}; hereafter H06) used Spitzer/MIPS to
measure the 24\,\micron\ system flux at five epochs over one orbit and
reported a finite amplitude phase curve consistent with zero phase
offset, though here we report a new analysis and interpretation of our
H06 data based on a better understanding of MIPS systematics.  The
8\,\micron\ observations of \cite{cowan:2007} also found variations
with zero phase offset for HD~179949b.  Observations of the less
intensely irradiated planet \hdoneb\ at 8\,\micron~and 24\,\micron\
\citep{knutson:2007b,knutson:2009a} revealed a relatively small
temperature contrast between the planet's day and night sides and a
$30\degg-40$\degg\ phase offset, indicating a moderate level of
eastward heat redistribution from the warm dayside to the cool night
side. It is important to note a possible observational bias: the first
two phase curves were sparsely sampled and are for non-transiting
systems with unknown orbital inclinations.  Though simulations suggest
inclination should not substantially affect a planet's observed phase
offset, the flux amplitude will be directly affected by inclination
\citep{rauscher:2008}; furthermore, phase curve interpretations are
more ambiguous without the absolute calibration provided by a
secondary eclipse \citep{burrows:2008}.

Several groups have hypothesized a connection between temperature
inversions and the magnitude of a system's phase offset
\citep[e.g.,][]{burrows:2008,fortney:2008}.  The favored (though
unproven) cause of inversions is a species residing at high altitude
that absorbs optically but is transparent to infrared radiation.  In
this scenario an inverted atmosphere absorbs stellar energy at lower
pressures where it should quickly reradiate to space; in a
non-inverted atmosphere the energy is absorbed much deeper, where it
may circulate farther around the planet and cause a measurable phase
offset.  \cite{showman:2009} do a fair job of reproducing the \hdoneb\
phase curves, but they predict secondary eclipse depths for the more
highly irradiated \hdtwob\ that do not match the observations of
\cite{knutson:2008}; \cite{burrows:2010} also model \hdtwob\ and they,
too, do not match the observed eclipse depths especially well. Thus,
our current understanding of the atmospheric structure and dynamics of
even the best-characterized planets still appears to be incomplete.

This is the context in which we obtained the high-cadence 24\,\micron\
phase curve of \uab\ described below.  We introduce the \uand\
system, and discuss our observations and data analysis, in \sref{obs}.
In \sref{results} we describe the planetary temperature contrast and
heat redistribution implied by our analysis.  We discuss possible
interpretations of our results in \sref{interpret}, and conclude in
\sref{conclusion}.

\section{Observations and Analysis}
\slab{obs}
\subsection{The  \ensuremath{\upsilon}~Andromedae System}
The planet \uab\ was one of the earliest reported hot Jupiters
\citep{butler:1997}, and the three-planet \ua\ system has been
observed numerous times in the years since
\citep{butler:2006,naef:2004,wittenmyer:2007,mcarthur:2010}.  The host
star has a spectroscopic effective temperature of $6212\pm64$~K
\citep{santos:2004} and a directly-measured diameter of
$1.631\pm0.014$~\rsun\ \citep{baines:2008}.  Spectroscopic and
isochronal mass estimates generally agree on a mass of
$\sim$1.3~\msun\ \citep{fuhrmann:1998, ford:1999, valenti:2005,
  takeda:2007}.  Because this system's planets do not transit we do
not know their physical sizes; however, if we assume \uab\ is a
typical hot Jupiter we can estimate its radius to be
$\sim1.3$~\rjup\footnote{Taken from the Extrasolar Planets
  Encyclopedia at \url{http://exoplanet.eu}}.

Using a combination of radial velocity and astrometry
\cite{mcarthur:2010} recently determined the orbits of the second and
third planets in the \ua\ system to be mutually inclined by 30\degg
. They suggest this nonplanar system may result from planet-planet
scattering that could also have moved the innermost planet, \uab ,
into its current orbit at 0.059~AU. The small stellar reflex motion
induced by \uab\ precluded a direct measurement of its orbital
inclination, but their preliminary numerical simulations extending
$10^5$~yr suggest \uab 's inclination may lie in the range
$\sim$20-45\degg\ (implying a planetary mass of 2-3\,\mjup ).  Though
their simulations did not fully explore the available parameter space,
the inclination range \cite{mcarthur:2010} suggest for \uab\ is
broadly consistent with the constraints we place on its inclination in
\sref{results}.

\subsection{Observations}
We observed the \ua\ system with Spitzer's MIPS 24\,\micron\ channel
\citep{rieke:2004} with observations spread across 1.2 orbits of \uab\
($\sim$5 days) during February 2009.  The observations consist of
seven brief ($\sim$3000 seconds on target) observational epochs and
one long, near-continuous observation $\sim$28~hours in length and
centered at phase 0.5 (secondary eclipse for transiting systems in
circular orbits, and the predicted time of flux maximum based on
H06). Our integrations total 18.5~hours.  Spitzer breaks up
observations into blocks of time called astronomical observation
requests (AORs) for instrument scheduling purposes: our short
observations consist of three sequential AORs, and the long observing
sequence consists of 71 AORs.  Altogether our data consist of 25\,488
frames, each with an integration time of 1.57 seconds.  We also observe
a flux calibrator star, HD~9712, in three two-AOR epochs, for a total
of 1.3~hours of integration.  The observations of \ua\ by H06, which
we reanalyze, consist of five AORs spaced over $\sim$5~days.

\subsection{Data Reduction}
We use the basic calibrated data (BCD) files generated by version
18.14 of the MIPS data reduction pipeline \citep{masci:2005}.  During
MIPS observations the instrument and spacecraft dither the target star
between fourteen positions on the detector
\citep[][Section~8.2.1.2]{SSC:2007}.  As noted previously
\citep[Deming et al.~2005; H06;][]{knutson:2009a} the MIPS detector
response is spatially nonuniform and so we treat the observations as
consisting of fourteen separate time series, modeling their
systematics separately in the final fit.  In each frame we measure
the system flux and the position of the star on the detector by fitting a
100$\times$ supersampled model MIPS PSF\footnote{Generated using Tiny
  Tim; available at \url{http://ssc.spitzer.caltech.edu/}} generated
using a 6200~K blackbody spectrum. We shift and scale the model PSF to
determine the best-fitting combination of background, stellar flux,
and PSF position.  Using position-dependent model PSFs does not
significantly change our results, so in all our photometry we use a
single model PSF generated at the center of the MIPS field of view.
We exclude hot pixels from the PSF fit by setting to zero the weight
of any frame's pixel that is more than 5$\sigma$ discrepant from the
median value of that pixel for all frames taken at that particular
dither position, and also exclude bad pixels flagged by the MIPS
reduction pipeline.

In each set of $\sim$40 frames, the first frame has $\sim$3\% higher
stellar flux, and the first several frames have a lower background,
than the rest of the frames in the set. These effects may be related
to the MIPS ``first-frame'' chip reset effect, though the effect we
see is qualitatively different from the one described in the MIPS
Instrument Handbook \citep{SSC:2010}.  We exclude the first frame in
each set from the remainder of our analysis, removing 708~frames. We
further exclude the first three contiguous AORs (700 frames, $\sim 40$
minutes), which are markedly discrepant from the final time series.
These first data were taken soon after a thermal anneal of the MIPS
detector, and during these observations we see anomalous readings in
the 24\,\micron\ detector anneal current, the scan mirror temperature,
and the MIPS B side temperature sensor.

The initial extracted photometry reveals a clear sinusoidal flux
variation, but with additional variability correlated with the
subpixel motion of the star on the detector, as shown in
Figures~\ref{fig:fluxcorrection} and~\ref{fig:rawflux}.  This effect
is likely from imperfect flat fielding rather than an intrapixel
sensitivity variation as seen in the Infrared Array Camera's 3.6 and
4.5\,\micron\ channels \citep{reach:2005,charbonneau:2005}, because
several time series with similar intrapixel locations exhibit
anticorrelated flux variations.  We tried creating a flat field by
median-stacking all the BCD frames after masking the region containing
the target star, but applying this flat field to the data before
computing photometry did not reduce the amplitude of the
position-correlated photometric variability.  As we describe below, we
suspect intermittent, low levels of scattered light interfere with our
ability to construct a sufficiently accurate flat field

Instead, we treat this variability by removing a linear function of
position from the computed photometry at each of the fourteen dither
positions.  Fitting functions with a higher-order dependence on
position, or including cross terms, does not change our final results
and increases the Bayesian Information Criterion ($\textrm{BIC}=\chi^2
+ k \ln N$, where $k$ is the number of free parameters and $N$ is the
number of points).  We remove the systematic effects, and assess
possible phase functions, by fitting the $j^{\textrm{th}}$ data point
at the $i^{\textrm{th}}$ dither position ($f_{ij}$) with the following
equation:
\begin{equation}
  f_{ij} = \left( a + b \sin \left[\Omega_{\textrm{orb}} t - \Phi_0 \right] \right) 
  \left( 1 + c_i + d_i x_{ij} + e_i y_{ij} \right)
\elab{model}
\end{equation}

This equation contains three astrophysical parameters and 42
instrumental parameters that account for the nonuniform detector
response. The astrophysical parameters of interest are the average
system flux $a$ and a time ($t$)-dependent sinusoidal phase function
with known planetary orbital frequency $\Omega_{\textrm{orb}}$ but
unknown amplitude $b$ and phase offset $\Phi_0$.  The remaining
parameters represent systematic effects to be removed: residual
sensitivity corrections $c_i$ and linear dependence on detector
position $(d_i,e_i)$ for each of the fourteen time series. 

To prevent parameters $a$ and $b$ from floating we artificially set
$c_1$ to satisfy the relation $\Pi_i (1 + c_i)=1$.  Ignoring the
$(d_i,e_i)$ factors does not change our final result for the 2009
dataset but increases the BIC, indicating a poorer fit to the data.
For the purposes of backwards comparison, we also apply our analysis
to the observations of H06.  Due to the limited temporal coverage of
this dataset including the $(d_i,e_i)$ does not improve the fit;
therefore we use only 17 parameters when reanalyzing this earlier
dataset. We otherwise apply the same data reduction steps as described
above.

In this work we consider only a sinusoidal phase function. Although we
recognize that the phase curve will not be truly sinusoidal in shape,
such a model is simple to work with and has a straightforward
interpretation as a two-hemisphere ``orange slice'' model \citep[][see also \sref{hemi} below]{cowan:2007,cowan:2008}.  Since the
motion-correlated flux variation only depends on relative motion
(rather than on absolute detector position) we normalized the
$(x_{ij}, y_{ij})$ in \eref{model} by subtracting the mean position in
each of the fourteen time series.

\subsection{Calibration and instrument stability}
Our continuous photometry reveals that the MIPS background flux
changes discontinuously from one AOR to the next.  Stellar photometry
is not similarly affected.  The background flux varies at the level
of 0.1\%, comparable in amplitude to the expected planetary signal.
We also see these discontinuities in background flux during MIPS
observations of \hdoneb\ \citep{knutson:2009a} and \hdtwob\
(unpublished; Spitzer Program ID 40280). 

Because the measured background level varies with the Spitzer AORs it
is extremely unlikely that this variability is of astrophysical
origin.  It is also unlikely that the background variability results
from the calibration process because we see the same effect in both
the raw and calibrated data products.  We observe no correlation
between the background variability and the various reported
instrumental parameters, though we cannot rule out either intermittent
scatter from other sources or slight changes in the detector bias.  A
global sensitivity drift does not seem to be the culprit because the
changes in background flux are uncorrelated with stellar photometry.

Although we were unable to determine the cause of the background
variations, we suspect that they are due to small changes in the
amount of scattered light reaching the MIPS detector.  Variable
scattered light could also explain our inability to remove the
motion-correlated photometric jitter with an empirical flat
field. Stellar photometry was substantially more stable on short
timescales than was the background in all of the extended MIPS
24\micron\ observations we examined, so we use this photometry in our
subsequent analysis.

Our observations include a flux calibrator star to check MIPS's
long-term photometric stability. MIPS 24\,\micron\ photometry is known
to be stable at the 0.4\% level over several years
\citep{engelbracht:2007}; if it were this variable on short timescales
we would be unable to discern the expected planetary emission.  We
observed the K1~III star HD~9712, taken from a catalog of bright
interferometric calibrator stars \citep{merand:2005}, during six AORs.
In our reduction of these data we do not apply a correction for pixel
motions and we achieve a repeatability of $\sim 10^{-4}$.  These
observations, shown as the red triangles in \fref{phasecurve}, imply
that the MIPS 24\,\micron\ detector was stable over the course of our
observations, and so we rule out detector sensitivity drifts as the
source of the observed flux variations.

The level of MIPS photometric stability was not well known early in
the Spitzer mission, so H06 used the background, attributed to
smoothly-varying zodiacal emission, as a calibrator to adjust the
photometry (see H06 Figure 1B).  As we now suspect the MIPS background
variations to result from scattered light in the instrument, there is
no longer any need, nor possibility, to use the zodiacal light as a
calibrator.  The H06 data were part of a preliminary Spitzer program
to assess variability of several systems for subsequent study in
programs such as ours.  Its five \ua\ AORs were taken at separate
epochs and there was no calibrator star, so it was impossible to make
the assessment described above.  We have reanalyzed the H06 data using
our new procedure and find values consistent with those plotted in H06
Figure 1A (see our \fref{phasecurve}) and also with our new data.
Thus, H06 did still present the first orbital phase variation for an
exoplanet.  The phase curve fit here to the H06 data provides accurate
parameters for that dataset.

\section{Results}
\slab{results} 

\subsection{System flux}
\slab{phot} Although our primary science result -- the planetary phase
curve described below -- is inherently a relative measurement, our
observations also allow us to measure precise, absolute 24\micron\
photometry for the \ua\ system.  We measure $F_\nu=0.488$~Jy and
$0.490$~Jy for the 2009 and 2006 datasets, respectively.  These fluxes
differ by 0.4\%, which is at the limit of the MIPS 24\micron\
precision; we therefore report the mean system flux as
$0.489\pm0.002$~Jy. This value is significantly discrepant from the
IRAS 25\micron\ flux of $0.73\pm0.05$~Jy \citep{moshir:1989}, but it
is consistent with a \cite{kurucz:1979} stellar spectrum tied to
optical and near-infrared photometry of \ua\ from the Tycho-2
\citep{hog:2000} and 2MASS \citep{skrutskie:2006} catalogs.

\subsection{A two-hemisphere model}
\slab{hemi}
As shown in \fref{phasecurve} and discussed below, we detect a flux
variation consistent with the orbital period of \uab .  This
measurement allows us to put tight constraints on the temperature
contrast and phase offset of the planet.  We interpret the observed
flux variation in the context of a planet composed of two blackbody
hemispheres of constant temperature -- i.e., a two-wedge ``orange
slice'' model \citep[c.f. ][]{cowan:2007}.  Sufficiently precise
observations of such a bifurcated planet will reveal a flux variation
with peak-to-trough amplitude
\begin{equation}
  \frac{\Delta F_P}{\langle F \rangle }  =  \frac{B_{\nu}(\tone) - B_{\nu}(\ttwo) }{B_{\nu}(\gamma T_{\textrm{eff}})} \left( \frac{R_P}{R_*} \right)^2 \sin i
  \textrm{,}
\elab{vconstraint}
\end{equation}
which normalizes the full amplitude of the planetary flux variation
$\Delta F_P$ by the mean flux from the system $\langle F \rangle$; in
\eref{model} $\Delta F_P / \langle F \rangle = 2b/a$. The quantity
$\gamma$ accounts for the star being fainter in the mid-infrared than
a blackbody with temperature $T_{\textrm{eff}}$; at 24\,\micron\ we
set $\gamma=0.8$ based on the models of \cite{kurucz:1979}. Thus,
measuring $\Delta F_P/\langle F \rangle$ gives the hemispheres'
brightness temperature contrast relative to the stellar flux, modulo a
$\sin i$ ambiguity.  By assuming a planetary albedo $A_B$ and assuming
all emitted radiation is reprocessed starlight, a second (bolometric)
relation obtains:
\begin{equation}
  \left( 1 - A_B \right) \frac{R_*^2}{2 a^2} T_{\textrm{eff}}^4 = \tone ^4 + \ttwo ^4 
  \textrm{.}
\elab{econstraint}
\end{equation} 

Thus, we implicitly assume that each hemisphere of our model planet
emits a bolometric flux equal to that of a blackbody with the
24\,\micron\ brightness temperature in \eref{vconstraint}.  Subject to
this assumption and given known or assumed values for the albedo $A_B$
and planetary radius $R_P$, one can use \erefs{vconstraint}{ and
}{econstraint} to determine the hemispheric temperatures \tone\ and
\ttwo\ at arbitrary orbital inclinations.  

Setting $\ttwo=0$ and $\tone=T_{P1,\textrm{max}}$ gives the minimum
planetary radius capable of reproducing the observed flux variation,
$\Delta F_P/\langle F \rangle$, as a function of the orbital
inclination.  Because this relation depends on inclination as $\left(
  \sin i \right)^{-1/2}$, measuring $\Delta F_P/\langle F \rangle$
gives an upper limit to the planet's surface gravity:
\begin{equation}
g \le \frac{G (m \sin i )}{R_*^2} 
  \frac{B_{\nu}(T_{P1,\textrm{max}})}{B_{\nu}(\gamma T_{\textrm{eff}})}
  \left( \frac{\Delta F_P}{\langle F \rangle} \right)^{-1}
  \textrm{.}
    \elab{maxgrav}
\end{equation}

\subsection{Model fits}
We determine the best-fit parameters using \citeauthor{powell:1964}'s
\citeyearpar{powell:1964} method for multivariate minimization (the
SciPy function optimize.fmin\_powell), and assess their uncertainties
and correlations with a Metropolis-Hastings, Markov-chain Monte Carlo
analysis \citep[MCMC; see][, section 15.8]{press:2007}.
\tref{coefs2009} (for the 2009 data) and \tref{coefs2006} (for the
2006 data) report these results.  \tref{param} lists the astrophysical
parameters of interest. We list the $\chi^2$ and BIC
values for the fits in \tref{stats}.  Parameter uncertainties are
estimated from distributions generated using the kernel density method
(KDE, implemented using the SciPy function stats.gaussian\_kde) by
determining the parameter values with equal KDE frequency that enclose
68\% of the distribution.

MCMC analysis evolves an initial set of parameters in a way that is
ultimately representative of their underlying probability
distributions.  For our Markov chain we choose a step size to give
approximately a 30\% step acceptance rate.  To adequately sample the
full parameter space we found it necessary to run the Markov chains
longer for the 2006 dataset than for the 2009 dataset. For the 2009
dataset we first ran the chain for $10^6$ burn-in steps and discarded
these; we then ran the chain for $2\times10^7$ steps, saving every
1000$^\textrm{th}$ step.  For the 2006 dataset our procedure is the
same but the burn-in phase lasted for $10^7$ steps and the chain was
then run for $5\times10^7$ steps, saving every 1000$^\textrm{th}$
step.  We inspected correlation plots for all possible parameter pairs
in both analyses to ensure adequate coverage of phase space and to
assess parameter correlations.

All the one-dimensional parameter distributions are unimodal and
approximately Gaussian in shape.  We see some correlations between the
$d_i$ and $e_i$, which is unsurprising given the degree of correlation
between the X and Y components of motion as shown in the middle panels
of \fref{fluxcorrection}.  More surprising is a correlation between
the mean system flux, $a$, and the phase offset, $\Phi_0$, as shown in
\fref{corr}.  Using a simulated dataset with white noise, we confirmed
that when forcing a fit to a sinusoid of known period, a slightly
higher mean value be counteracted by a slightly lower phase offset;
however, we observe the opposite correlation. In any case no
significant correlation is apparent between the phase curve amplitude
$\Delta F_P / \langle F \rangle$ and the phase offset, which are the
primary quantities of interest for our analysis. The best $\chi^2$
from the MCMC is consistent to within a small fraction of the
uncertainties with the optimizer values.

The model in \eref{model} provides a good fit to both the
astrophysical flux modulation and the instrumental flux variations at
each of the fourteen dither positions, as shown in \fref{rawflux}. We
plot the photometry after removal of the systematic effects in
\fref{phasecurve}, along with the best-fitting sinusoidal phase curve,
for both the 2006 and 2009 datasets. Both datasets appear to vary
approximately in phase; this coherence is a strong argument that the
flux modulation we see is due to the planet \uab .  The
goodness-of-fit statistics ($\chi^2$ and BIC) for both datasets are
listed in Table~\ref{tab:stats}.  The high $\chi^2$ for the 2009
dataset probably results from a somewhat non-sinusoidal phase curve
and from the residual systematics apparent in \fref{phasecurve}.

\subsection{Phase curve amplitude}
\slab{temp} We measure values of $\Delta F_P/\langle F \rangle$ of
$\Delta F_P/\langle F \rangle = 0.001300\pm0.000074$ for the 2009 data
and $0.00090 \pm 0.00022$ for our reanalysis of the 2006 data, as
shown in \tref{param}.  Using the absolute calibration from
\sref{phot}, we find absolute peak-to-trough phase curve amplitudes of
$0.636\pm0.036$~mJy and $0.44\pm0.11$~mJy for the 2009 and 2006
datasets, respectively. Thus the detection of the phase curve
amplitude at both epochs is statistically significant at the
$>4\sigma$ level and is substantially smaller than reported by H06
(due to the calibration issues discussed above).  The two epochs'
phase curve amplitudes are consistent at the 1.7$\sigma$ level; thus
there is no evidence that the planetary emission exhibits inter-epoch
variability.  The lack of variability is consistent with the recent
results of \cite{agol:2010}, who set an upper limit of 2.7\% on
\hdoneb 's dayside flux variations.

As \uab 's orbit is inclined toward face-on, a greater intrinsic
temperature contrast is required to generate the observed flux
variation.  Using \eref{vconstraint} and our measurement of the phase
curve amplitude we determine the expected day/night contrast ratio and
plot it in the upper panel of \fref{contrast}. We also plot the upper
limits on the day/night contrast assuming planetary radii of
1.3~(1.8)~\rjup ; the implication is that the planet's orbital
inclination angle is likely $\gtrsim 28\degg$ ($14\degg$).  These
limits complement the preliminary limits on \uab 's orbital
inclination from the stability modeling of \cite{mcarthur:2010}, which
suggest $i\sim20\degg-45\degg$.

Invoking \eref{econstraint} allows us to determine the brightness
temperatures of the planetary hemispheres in our model at each
inclination angle. We assume zero albedo \citep[c.f.][]{rowe:2008} and
a planetary radius of 1.3~\rjup\ and plot the hemispheres'
temperatures and 3$\sigma$ limits in the lower panel of
\fref{contrast}.  This sets a lower bound to the temperature contrast
between the two hemispheres to be $\tone - \ttwo \gtrsim 900$~K.  The
hotter hemisphere's temperature remains in the range $\sim$1700-1900~K
as we vary the radius from 1.0~\rjup\ to 1.8~\rjup ; however larger
radii result in higher temperatures for the cooler hemisphere to
maintain the measured flux and thus decrease the temperature contrast.

Using \eref{maxgrav} we find that \uab 's surface gravity is
$<2100$~cm~s$^{-2}$ with 3$\sigma$ confidence: this result is
independent of assumptions about the planet's radius or orbital
inclination. For a hot hemisphere temperature of $\sim$1800~K
(c.f. \fref{contrast}) this limit on the surface gravity implies an
atmospheric scale height $>300$~km.  In \fref{mr} we plot the allowed
regions of mass-radius parameter space against the known population of
transiting extrasolar planets.  Thus, our measurements suggest that
\uab\ has a lower surface gravity than Jupiter, \hdoneb , and a number
of other transiting extrasolar planets. This result demonstrates that
\uab\ is indeed a gaseous Jovian planet, but we cannot determine
whether it is a highly inflated planet or whether it is dominated by a
sizeable rocky core.

\subsection{Phase offset}
\slab{offset}
Because \uab\ does not transit its host star we know its orbital
ephemeris less precisely than we do for transiting planets. We used
the Systemic Console\footnote{available from
  \url{http://oklo.org/downloadable-console/}} \citep{meschiari:2009}
to reanalyze the published radial velocity data of \ua\
\citep{butler:1997,naef:2004,butler:2006,wittenmyer:2007,mcarthur:2010}
using Systemic's Levenberg-Marquardt algorithm and ignoring system
stability constraints.  We obtain orbital parameters consistent with
those of \cite{mcarthur:2010}.  By providing the covariances between
the various fit parameters, this reanalysis allows a substantially
more precise estimate of the planetary ephemeris than is available
from the literature.  We compute a time of zero relative radial
velocity (phase 0.5, or secondary eclipse in a circular transiting
system) of JD\,=\,2\,454\,868$.78\pm0.07$ ($1\sigma$), which
corresponds to an uncertainty of $\sim6\degg$ in determining the phase
offset.

Assuming a two-hemisphere model, we find a phase offset of $84.5\degg
\pm 2.3\degg$ relative to the computed ephemeris.  This uncertainty
may be an underestimate since we artificially constrain our phase
curve to be sinusoidal.  \cite{knutson:2009a} discuss the artificially
low uncertainties obtained from fitting to an arbitrarily chosen
model, though here we have substantially broader phase coverage than
was available in that study.  The system flux reaches a maximum before
phase 0.5, indicating that the brighter hemisphere is offset to the
east of the substellar point, as observed for \hdoneb\
\citep{knutson:2007b,knutson:2009a,agol:2010}.

This large phase offset is strikingly different from the near-zero
phase offset reported by H06; this difference is due to the choice of
system calibration as discussed above. From our reanalysis of the 2006
data we find a phase offset of $57\degg \pm 21\degg$ relative to our
ephemeris.  The phase offsets at the two epochs are consistent at the
1.3$\sigma$ level, and thus there is no evidence for inter-epoch
variability in the phase offset.

\section{Discussion}
\slab{interpret} The most striking result of our analysis is the large
phase offset evident in the light curve. The direction of the phase
offset is broadly consistent with the prediction by many circulation
models of a large-scale, high-velocity, eastward-flowing jet on hot
Jupiters \citep{cho:2003, cooper:2005,cho:2008,
  showman:2009,burrows:2010,rauscher:2010,thrastarson:2010} and as
seen on \hdoneb\ \citep{knutson:2007b,knutson:2009a}. However, the
magnitude of the phase offset is far larger than is predicted at the
low pressures characteristic of the 24\,\micron\ photosphere
\citep{showman:2009}. 

A partial explanation for such a large phase offset could be that \uab
's atmosphere is substantially transparent to the incident stellar
flux, with the result that the insolation is deposited at sufficient
depth for substantial advection to occur.  In this case, the planet
might lack an atmospheric temperature inversion. Indeed, the large
phase offset is more consistent with expectations for a planet lacking
a temperature inversion than with expectations for a planet with an
inverted atmosphere \citep[c.f. \hdoneb\
vs. \hdtwob;][]{showman:2009}.  \cite{cooper:2005,cooper:2006}
predicted maximum hemisphere-averaged temperatures to be offset by
$\gtrsim 60\degg$ for an irradiated planet with no high-altitude
absorbers, but it is unlikely that this would translate into high
phase offsets at 24\,\micron\ due to the high altitude of the
photosphere at this wavelength. The absence of a temperature inversion
is at odds with the recent proposal by \cite{knutson:2010} that, due
to the lack of activity apparent in optical spectra of \ua one would
expect \uab\ to have a temperature inversion.  \cite{thrastarson:2010}
also show clear shifted hot regions at higher pressure; the deep
vortices produced by their simulations are several hundreds of degrees
hotter than the surroundings.

Setting aside the phase offset for a moment, the phase curve amplitude
we measure could be directly interpreted in the context of the models
of \cite{showman:2009} as a typical phase curve for a planet with
orbital inclination of 50\degg -60\degg . However, simulations with
greater phase offsets tend to show lower phase curve amplitudes
\citep[e.g.][]{showman:2009,burrows:2010}; thus it is difficult to
simultaneously explain both amplitude and offset and it is difficult
to reconcile current theory with our observations.

Alternatively, the large phase offset we see could represent
reradiation of thermal energy deposited in shock fronts in the
planet's atmosphere.  Many simulations predict supersonic equatorial
jets on hot Jupiters that carry substantial kinetic energy; shocks
could manifest themselves where the jet transitions to subsonic
speeds.  \cite{rauscher:2010} note that their simulations, as well as
those of \cite{showman:2009}, exhibit structures which could be
interpreted as shocks -- however, these models do not explicitly treat
shock physics. \cite{dobbs-dixon:2010} observe similar features in
their models using an additional artificial viscosity factor to
simulate shock behavior, and note that in high-altitude, high-velocity
regions the energy carried by kinetic energy becomes comparable to the
enthalpic energy.  \cite{watkins:2010} have recently suggested that
gravity waves in the atmosphere of a hot Jupiter can also heat the
planet's upper atmosphere. It is unclear whether either shocks or
gravity waves can deposit sufficient energy in the 24\,\micron\
photosphere to cause the large phase offset we see.  We look forward
to further research into these topics to determine whether these or
other phenomena can explain our observations.

We must also consider the possibility that the periodic flux
modulation we see is intrinsic to the star rather than emanating from
the planet.  \cite{shkolnik:2005,shkolnik:2008} report evidence for
intermittently periodic stellar activity in this system correlated
with \uab 's orbital period.  They interpret this periodicity as a
possible magnetospheric star-planet interaction, but they detect this
periodicity at only some of their observational epochs.  The
consistency between our analysis of the 2006~and~2009 MIPS data
suggests we are not seeing such a transient phenomenon, although
analysis of additional activity measurements (e.g., from the Keck
Observatory Archive) spanning the time of our Spitzer observations
would help solidify this claim.  \cite{shkolnik:2005} also report
variations consistent with zero phase offset, which disagree with our
observed phase offset.  In addition, 24\,\micron\ stellar variability
at the level we observe would imply much greater variability at
optical wavelengths: this is not observed
\citep{henry:2000activity}. Finally, as discussed by H06, energy
considerations indicate that such an intense star-planet interaction
would cause the planet to spiral into the star in $\lesssim
10^7$~yr. Thus, the 24\,\micron\ flux variations are likely to be of
planetary origin.

Although a general framework exists within which to interpret
observations of exoplanetary atmospheres, our understanding is still
extremely limited.  \cite{madhusudhan:2009} demonstrate that some
planets with claimed inversions can also be fit by non-inverted
atmospheric models due to much greater number of free model parameters
versus the limited number of observational constraints.  In addition,
there may be reason to question the reliability of some of the
circulation models currently in use.  \cite{thrastarson:2010}, in
their extensive exploration of initial condition parameter space, have
recently shown the extreme susceptibility of at least one circulation
model to minute variations in initial conditions, resulting in
substantial variability in final ``steady-state'' temperature
contrasts and phase offsets. This chaotic behavior calls into question
the ability of at least some models to make accurate, qualitative
predictions about any of the quantities we are interested in.  Other
models \citep[e.g.,][]{dobbs-dixon:2010} exclude planets' polar
regions from their simulations, where large vortices are often seen to
form \citep{cho:2003,cho:2008, rauscher:2008}.  While the polar
regions present a small cross-sectional area in transiting systems,
for non-transiting systems such as \uab\ polar emission will
constitute a larger component of the observed system flux.  A fully
consistent three-dimensional circulation geometry is essential for
comparison to our observations.  Nevertheless, it is important to
remember that general circulation models have trouble predicting
global weather patterns even for relatively well-studied solar system
planets; thus, a comprehensive, quantitative understanding of
extrasolar planetary atmospheric dynamics will likely remain elusive
for some time to come.

\section{Conclusions and Future Work}
\slab{conclusion} We have described a new 24\,\micron\ phase curve,
which we interpret as being due to emission from the planet \uandb\
modulated by the planet's orbit.  Using a simple two-hemisphere model
we determine the peak-to-trough phase curve amplitude to be $0.001300
\pm 0.000074$. This result suggests an average ``hot side'' temperature
of $\sim$1800~K; for an average-sized hot Jupiter (1.3\,\rjup) this
implies a hemisphere-averaged planetary temperature contrast of
$\sim$900~K and an orbital inclination $i\gtrsim28\degg$.

We find a phase offset of $84\degg \pm 2 \degg \pm 6 \degg$, where we
break the uncertainty into the error relative to our ephemeris and the
error in our ephemeris, respectively. Such a large phase offset is
difficult to reconcile with most current models. The phase curve is
hotter hemisphere is offset to the east, as previously observed for
\hdoneb\ \citep{knutson:2007b,knutson:2009a}.

We reanalyze our earlier (H06) observations of this system and find a
phase curve amplitude of $0.00090 \pm 0.00022$ and a phase offset of
$57\degg \pm 21 \degg \pm 6 \degg$.  There is no evidence for
inter-epoch variability in the planetary phase curve. This is
primarily due to the large uncertainties from the 2006 dataset and
demonstrates the difficulty in measuring such variability with
sparsely-sampled phase curves.

There are substantial challenges in interpreting a phase curve
observed at only a single wavelength due to the degeneracies between
planetary radius, orbital inclination, and atmospheric composition and
structure \citep{burrows:2008}.  Some of these difficulties could be
mitigated with phase curve measurements at additional wavelengths --
ideally from space (i.e., warm Spitzer) but also potentially from
ground-based near-infrared observations \citep[e.g.,][]{barnes:2010}.
For example, commensurate phase curve amplitudes at both 3.6\,\micron\
and 4.5\,\micron\ would suggest that \uab\ does indeed lack a
temperature inversion.  On the other hand, differing phase curve
amplitudes at these wavelengths could suggest an inversion and be more
difficult to reconcile with the large phase offset at 24\,\micron .
Phase offsets in these or other scenarios would also depend on the
particular atmospheric temperature structure of the planet. Phase
curves at 24\,\micron\ currently exist for only two planets, \uab\ and
\hdoneb\ (though unpublished data exist for \hdtwob ). Cool,
inversionless \hdoneb\ is probably the best-characterized extrasolar
planet, and additional phase curves are already being observed for
this object at 3.6\,\micron\ and 4.5\,\micron .  Phase curves of
additional planets at multiple wavelengths are essential to ensure
that our evolving views of the atmospheres of hot Jupiters are not
biased by unbalanced datasets.

Whatever the cause of the substantial energy transport implied by the
large phase offset we measure, if this phenomenon occurs in other
(transiting) systems there are important implications for transmission
spectroscopy.  Line-of-sight effects cause optical transmission
spectra to probe pressures comparable to those probed by mid-infrared
emission; temperatures near the planetary terminator of $\sim$1800~K
(as we observe) should easily be detectable with ground- or
space-based spectra.  \cite{sing:2008} and \cite{lecavelier:2008}
infer a terminator temperature on \hdtwob\ of $2200\pm260$~K at
$33\pm5$~mbar -- roughly at the expected 24\,\micron\ photosphere --
which is not dissimilar from the hot terminator-centered hemisphere we
observe on \uab . Further phase curve and transmision spectra of
additional systems are needed to determine whether this hot,
high-altitude terminator measurement results from a mechanism similar
to what we observe on \uab .

We note that \uab\ is too bright to be observed photometrically at
shorter wavelengths with the planned James Webb Space Telescope
(though spectroscopy may be feasible).  Given JWST's magnitude limits
and the expected high demand for its observing time, the community
should consider a dedicated space-based mid-infrared photometry and
spectroscopy mission \citep{vasisht:2008}. Such a mission would allow
uninterrupted long-term monitoring of nearby hot Jupiter systems.
This would provide high-precision measurements of these systems'
thermal emission and energy distributions, and possibly provide the
first definitive evidence of dynamical meteorological processes --
weather -- on extrasolar planets.

\acknowledgments
We thank J.~Colbert, C.~Engelbracht, and G.~Rieke for help in
interpreting MIPS systematics, and T. Loredo for helpful discussions
of statistics.  This work is based in part on observations made with
the Spitzer Space Telescope, which is operated by the Jet Propulsion
Laboratory, California Institute of Technology under a contract with
NASA.  Support for this work was provided by NASA through an award
issued by JPL/Caltech. We received free software and services from
SciPy, Matplotlib, and the Python Programming Language.  This research
made use of Tiny Tim/Spitzer, developed by John Krist for the Spitzer
Science Center; the Center is managed by the California Institute of
Technology under a contract with NASA. Part of this work was performed
while in residence at the Kavli Institute for Theoretical Physics,
funded by the NSF through grant number PHY05-51164.

\clearpage
\begin{deluxetable}{c rcl l}  
\tablecolumns{5}
\tablecaption{\label{tab:coefs2009} Phase curve and calibration parameters from \eref{model} (2009 data)\tablenotemark{a}. }
\startdata
$a$\tablenotemark{b}     &  0.488436&$\pm$&    0.000011  Jy &\\
$b     $        &       0.000317    &$\pm$   &       0.000017  Jy & \\
$\Phi_0\tablenotemark{c}$        &       84.5\degg       &$\pm$   &       2.3\degg &\\
$c_{1} $        &       0.005367    &$\pm$   &       0.000054 & \\
$c_{2} $        &       0.012926    &$\pm$   &       0.000055 & \\
$c_{3} $        &       0.001180    &$\pm$   &       0.000055 & \\
$c_{4} $        &       0.009498    &$\pm$   &       0.000055 & \\
$c_{5} $        &       0.001548    &$\pm$   &       0.000054 & \\
$c_{6} $        &       0.014334    &$\pm$   &       0.000027 & \\
$c_{7} $        &       -0.006333    &$\pm$   &       0.000054 & \\
$c_{8} $        &       -0.004193    &$\pm$   &       0.000054 & \\
$c_{9} $        &       -0.002129    &$\pm$   &       0.000054 & \\
$c_{10} $       &       -0.007009    &$\pm$   &       0.000055 & \\
$c_{11} $       &       -0.000549    &$\pm$   &       0.000055 & \\
$c_{12} $       &       -0.009458    &$\pm$   &       0.000054 & \\
$c_{13} $       &       -0.002757    &$\pm$   &       0.000054 & \\
$c_{14} $       &       -0.012002    &$\pm$   &       0.000054 & \\
$d_{1} $        &       -0.00429     &$\pm$  &       0.00072 & \\
$d_{2} $        &       0.00146     &$\pm$  &       0.00070 & \\
$d_{3} $        &       -0.00214     &$\pm$  &       0.00071 & \\
$d_{4} $        &       -0.00428     &$\pm$  &       0.00070 & \\
$d_{5} $        &       -0.00520     &$\pm$  &       0.00071 & \\
$d_{6} $        &       -0.00177     &$\pm$  &       0.00070 & \\
$d_{7} $        &       -0.00100     &$\pm$  &       0.00078 & \\
$d_{8} $        &       -0.00049     &$\pm$  &       0.00072 & \\
$d_{9} $        &       0.00196     &$\pm$  &       0.00066 & \\
$d_{10} $       &       -0.00252     &$\pm$  &       0.00067 & \\
$d_{11} $       &       -0.00029     &$\pm$  &       0.00066 & \\
$d_{12} $       &       -0.00409     &$\pm$  &       0.00067 & \\
$d_{13} $       &       0.00259     &$\pm$  &       0.00068 & \\
$d_{14} $       &       -0.00152     &$\pm$ &       0.00068 & \\
$e_{1} $        &       0.00066     &$\pm$  &       0.00080 & \\
$e_{2} $        &       0.00031     &$\pm$  &       0.00079 & \\
$e_{3} $        &       0.00211     &$\pm$  &       0.00040 & \\
$e_{4} $        &       0.00218     &$\pm$  &       0.00078 & \\
$e_{5} $        &       0.00409      &$\pm$ &       0.00080 & \\
$e_{6} $        &       0.00009     &$\pm$  &       0.00080 & \\
$e_{7} $        &       0.00211     &$\pm$  &       0.00081 & \\
$e_{8} $        &       -0.00259     &$\pm$  &       0.00076 & \\
$e_{9} $        &       0.00628     &$\pm$  &       0.00076 & \\
$e_{10} $       &       -0.00284     &$\pm$  &       0.00039 & \\
$e_{11} $       &       -0.00168     &$\pm$  &       0.00075 & \\
$e_{12} $       &       -0.0029     &$\pm$  &       0.0011 & \\
$e_{13} $       &       -0.00061     &$\pm$  &       0.00076 & \\
$e_{14} $       &       0.00025     &$\pm$  &       0.00074 & \\
\enddata                             
\tablenotetext{a}{ Errors quoted are the 68.3\% confidence limits.
  The first three parameters are the mean system flux $a$, the phase
  curve half-amplitude $b$, and the phase offset $\Phi_0$. The other
  coefficients represent fits to the nonuniform detector response at
  each of the fourteen MIPS dither positions.  The $c_i$ are
  dimensionless, and the $d_i$ and $e_i$ are in units of
  $\textrm{pixel}^{-1}$.}  \tablenotetext{b}{The absolute accuracy of
  the mean system flux $a$ is 0.4\% as discussed by
  \cite{engelbracht:2007}.}  \tablenotetext{c}{$\Phi_0$ is measured
  relative to our computed ephemeris, as discussed in \sref{offset}.}
\end{deluxetable}

\begin{deluxetable}{c r c l l}
\tablecolumns{5}
\tablecaption{\label{tab:coefs2006} Phase curve and calibration parameters from \eref{model} (2006 data)\tablenotemark{a}. }
\startdata
$a$\tablenotemark{b}   &   0.490250 &$\pm$  &   0.000039   Jy &\\
$b     $        &      0.000221     &$\pm$  &       0.000053   Jy &\\
$\Phi_0$\tablenotemark{c}        &       57\degg       &$\pm$  &       21\degg  & \\
$c_{1} $        &       0.00284     &$\pm$  &       0.00030  & \\
$c_{2} $        &       0.01423     &$\pm$  &       0.00028  & \\
$c_{3} $        &       0.00112     &$\pm$  &       0.00030  & \\
$c_{4} $        &       0.01094     &$\pm$  &       0.00028  & \\
$c_{5} $        &       0.00235     &$\pm$  &       0.00029  & \\
$c_{6} $        &       0.01574     &$\pm$  &       0.00028  & \\
$c_{7} $        &       -0.00757     &$\pm$  &       0.00027  & \\
$c_{8} $        &       -0.00596     &$\pm$  &       0.00030  & \\
$c_{9} $        &       -0.00086     &$\pm$  &       0.00030  & \\
$c_{10} $       &       -0.00755     &$\pm$  &       0.00029  & \\
$c_{11} $       &       -0.00097     &$\pm$  &       0.00029  & \\
$c_{12} $       &       -0.00945     &$\pm$  &       0.00029  & \\
$c_{13} $       &       -0.00299     &$\pm$  &       0.00027  & \\
$c_{14} $       &       -0.01138     &$\pm$  &       0.00027  & \\
\enddata
\tablenotetext{a}{Errors quoted are the 68.3\% confidence limits.
  The first three parameters are the mean system flux $a$, the phase
  curve half-amplitude $b$, and the phase offset $\Phi_0$. The other
  coefficients represent fits to the nonuniform detector response at
  each of the fourteen MIPS dither positions.  The $c_i$ are
  dimensionless, and the $d_i$ and $e_i$ are in units of
  $\textrm{pixel}^{-1}$.}  \tablenotetext{b}{The absolute accuracy of
  the mean system flux $a$ is 0.4\% as discussed by
  \cite{engelbracht:2007}.}  \tablenotetext{c}{$\Phi_0$ is measured
  relative to our computed ephemeris, as discussed in \sref{offset}.}
\end{deluxetable}

\begin{table}[h]
\caption{Astrophysical parameters of interest}
\tlab{param}
\begin{center}
\begin{tabular}{c | c c c | c c c }
 &         \multicolumn{3}{c|}{2006}      &      \multicolumn{3}{c}{2009}    \\
\hline
$\Delta F / F$  &       0.00090        &       $\pm$   &       0.00022      &       0.001300        &       $\pm$   &       0.000074\\
$\Phi_0$        &       57\degg   &       $\pm$   &       21\degg &       84.5\degg   &       $\pm$   &       2.3\degg \\
& & & & & & \\
\end{tabular} 
\end{center}
\end{table}

\begin{table}[h]
  \caption{Goodness-of-fit statistics}
\tlab{stats}
\begin{center}
\begin{tabular}{c c c}
                &       2006    &       2009 \\
$\chi^2$        &       814.1   &       38422.7 \\
$k$             &       17      &       45 \\
$N$             &       838     &       23884 \\
BIC             &       928.6   &       38876.4 \\
\end{tabular}
\end{center}
\end{table}    

\clearpage

\fig{fluxcorrection}{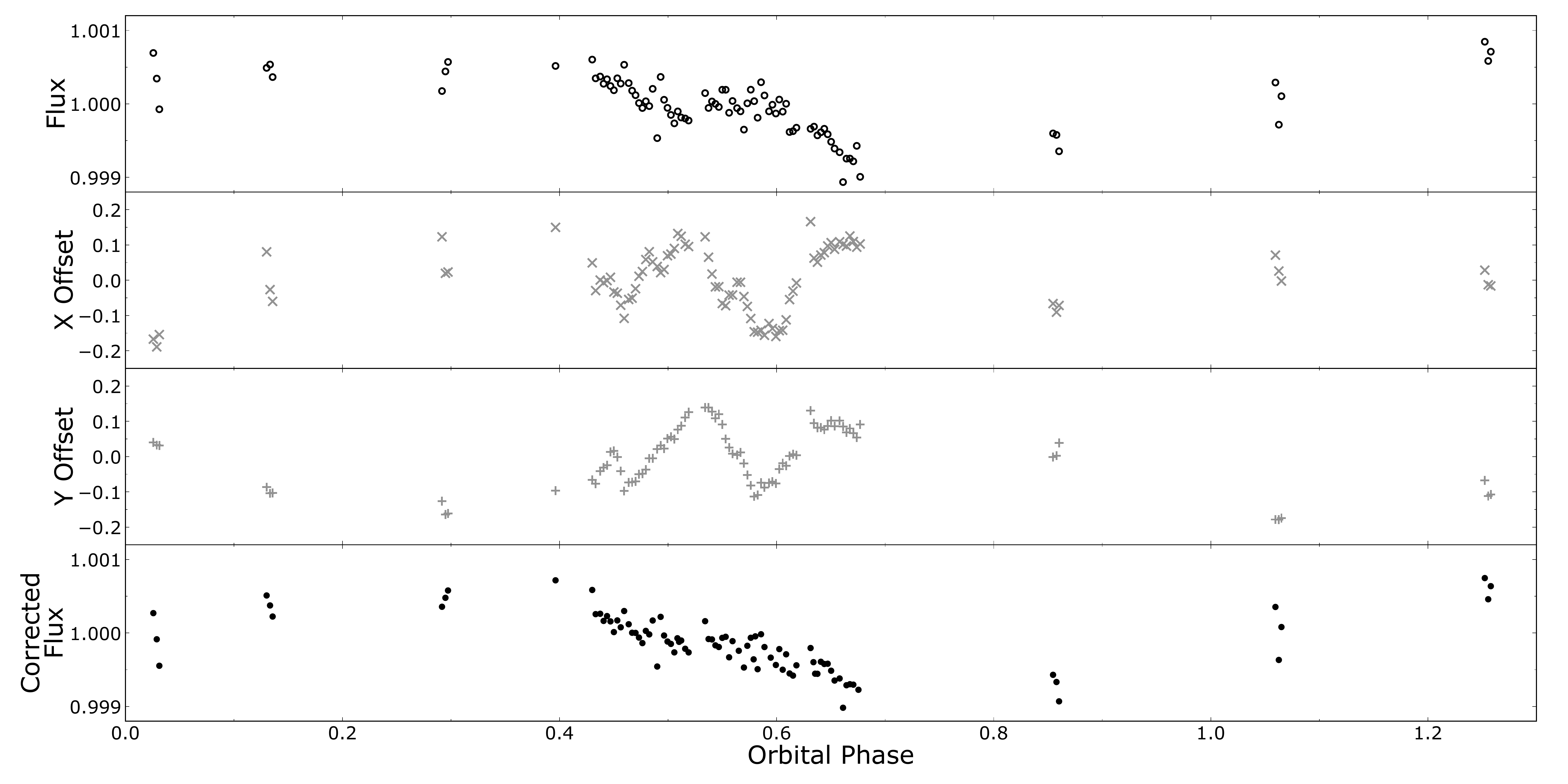}{width=7in}{}{Removal
  of pointing-correlated photometric variability.  The top panel shows
  photometry after bringing all fourteen individual time series to a
  common level.  A large-scale sinusoidal flux variation is evident,
  but so is a shorter-scale ``ripple'' (near phase 0.5); this ripple
  is correlated with the motion of \ua\ on the MIPS detector, plotted
  as X and Y in the middle two panels.  The bottom plot shows the
  final, cleaned photometry after removing this correlation. For
  display purposes the full dataset has been averaged over each AOR in
  this figure. Uncertainties on the X and Y points are typically
  $\sim$$10^{-3}$ pixel, while the uncertainties of the photometric
  points (not plotted for clarity) is typically $\sim$$1.5 \times
  10^{-4}$.}

\fig{rawflux}{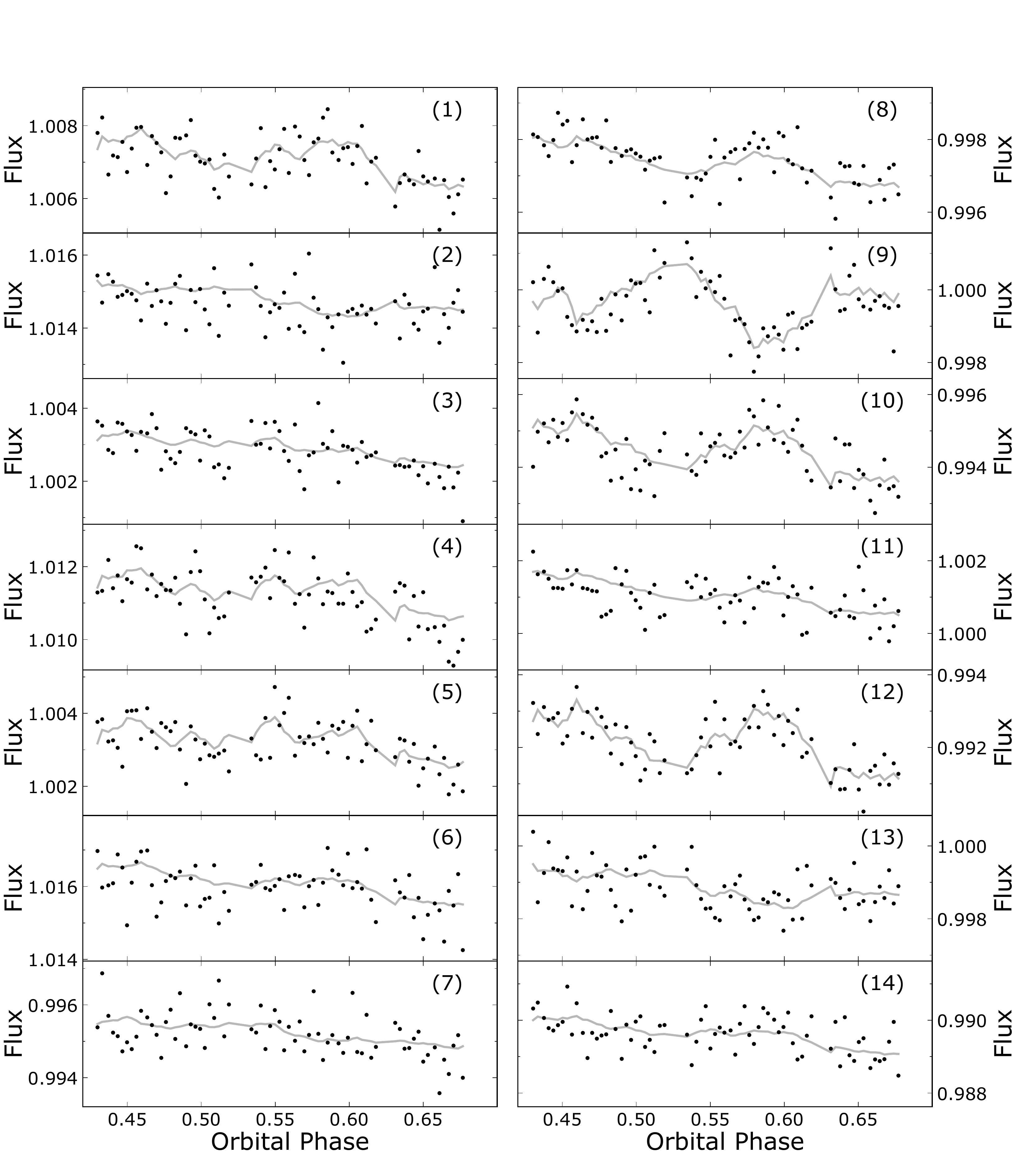}{width=6.2in}{}{Raw photometry (dots)
  and the best-fit model (solid lines) at each of the fourteen dither
  positions.  The measured flux varies by several percent from one
  position to another, as evidenced by the different scales in each
  panel. The downward trend evident in all panels is due to the
  decreasing flux from the system, shown more clearly in
  \fref{phasecurve}.  For display purposes the data have been averaged
  over each AOR and we plot only the continuous-observing segment of
  our observations. The precision of individual points (not plotted
  for clarity) is $\sim 5\times 10^{-4}$. }

\figtwocol{phasecurve}{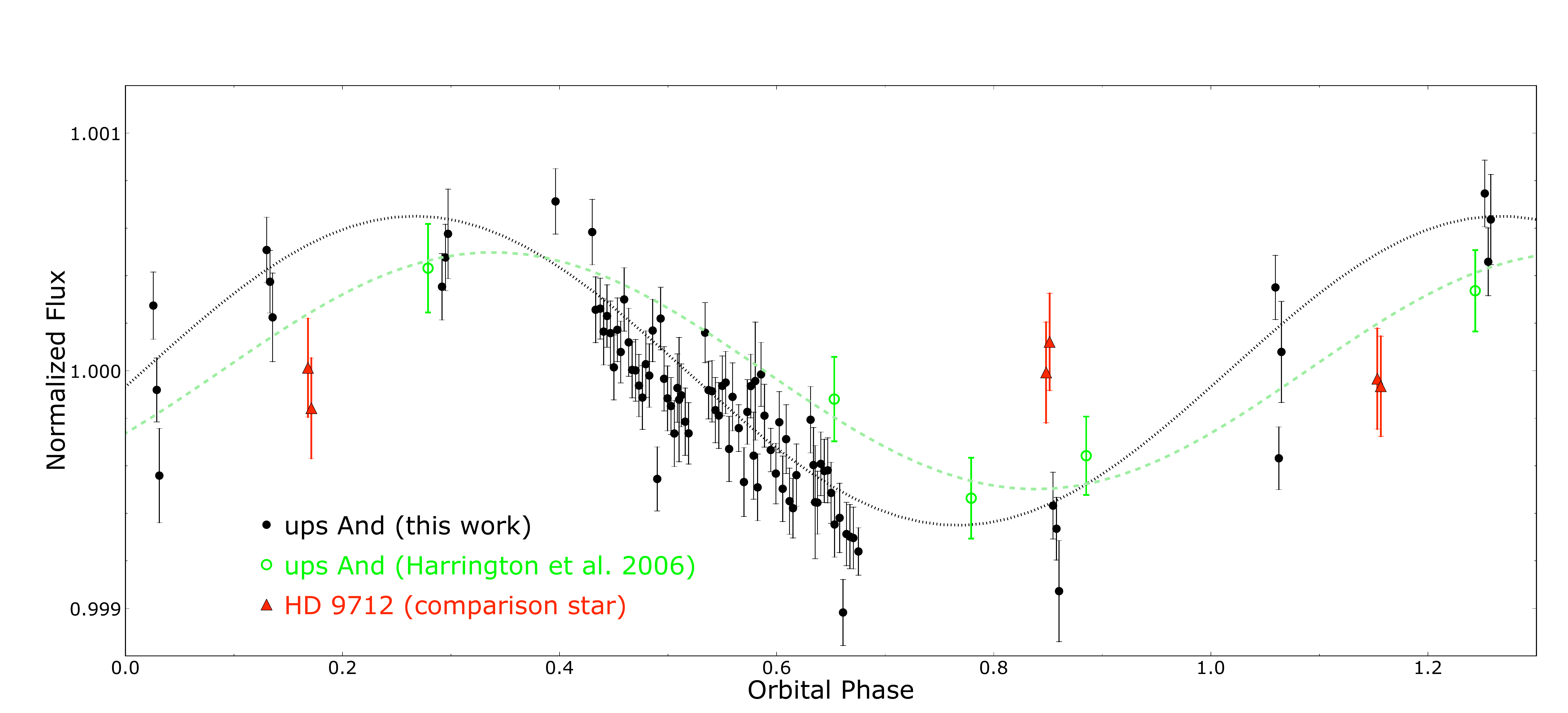}{width=7in}{}{Phase curve of
  the \ua\ system, phased to the orbit of the innermost planet (black
  circles).  Position-dependent sensitivity effects have been removed
  and for display purposes we have averaged the data over each
  AOR. The best-fit sinusoid (black dotted line) exhibits a phase
  offset of $\sim80$\degg, consistent with a planetary ``hot spot''
  advected almost to the planet's day-night terminator.  The light
  green circles are our reanalysis of the data from
  \protect{\cite{harrington:2006}}, and the light green dashed line
  shows the best-fit sinusoid to these data; the phase coherence
  between the 2006 and 2009 datasets is consistent with flux modulated
  by the innermost planet's orbit.  The red triangles show our flux
  calibrator observations, which are consistent with a constant
  detector sensitivity.}

\fig{corr}{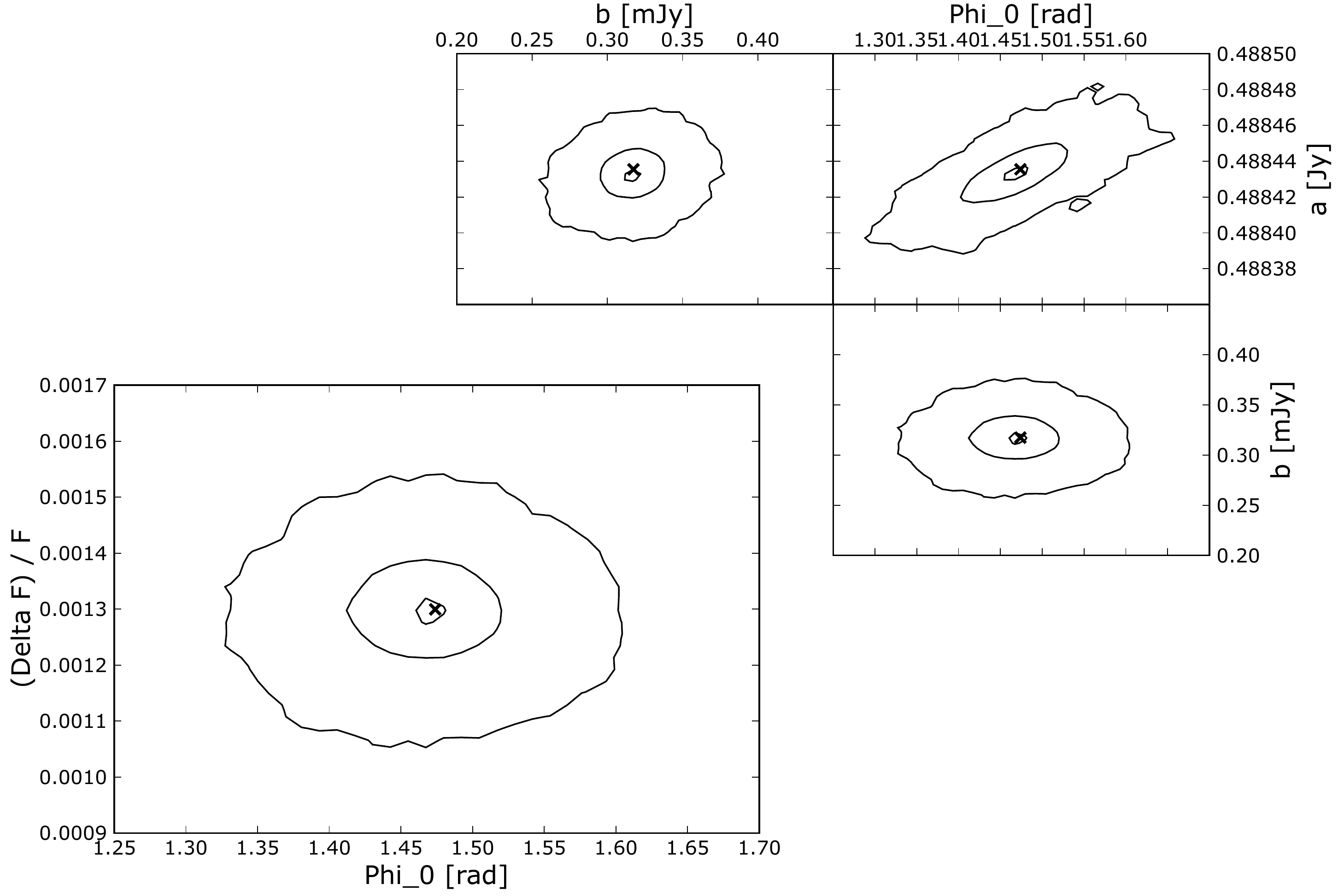}{width=7in}{}{Two-dimensional confidence
  intervals on the astrophysical parameters of interest in
  \eref{model}: the relative phase curve amplitude $\Delta F_P/F$, the
  absolute mean system flux $a$, the phase curve half-amplitude $b$,
  and the phase offset $\Phi_0$.  The solid lines are the contours
  that enclose 68.3\%, 95.4\%, and 99.7\%\ of the parameter space from
  the 2009 dataset.  The `$\times$' in each panel marks the best-fit
  parameter listed in Tables~\ref{tab:coefs2009} and~\ref{tab:param}.}

\fig{contrast}{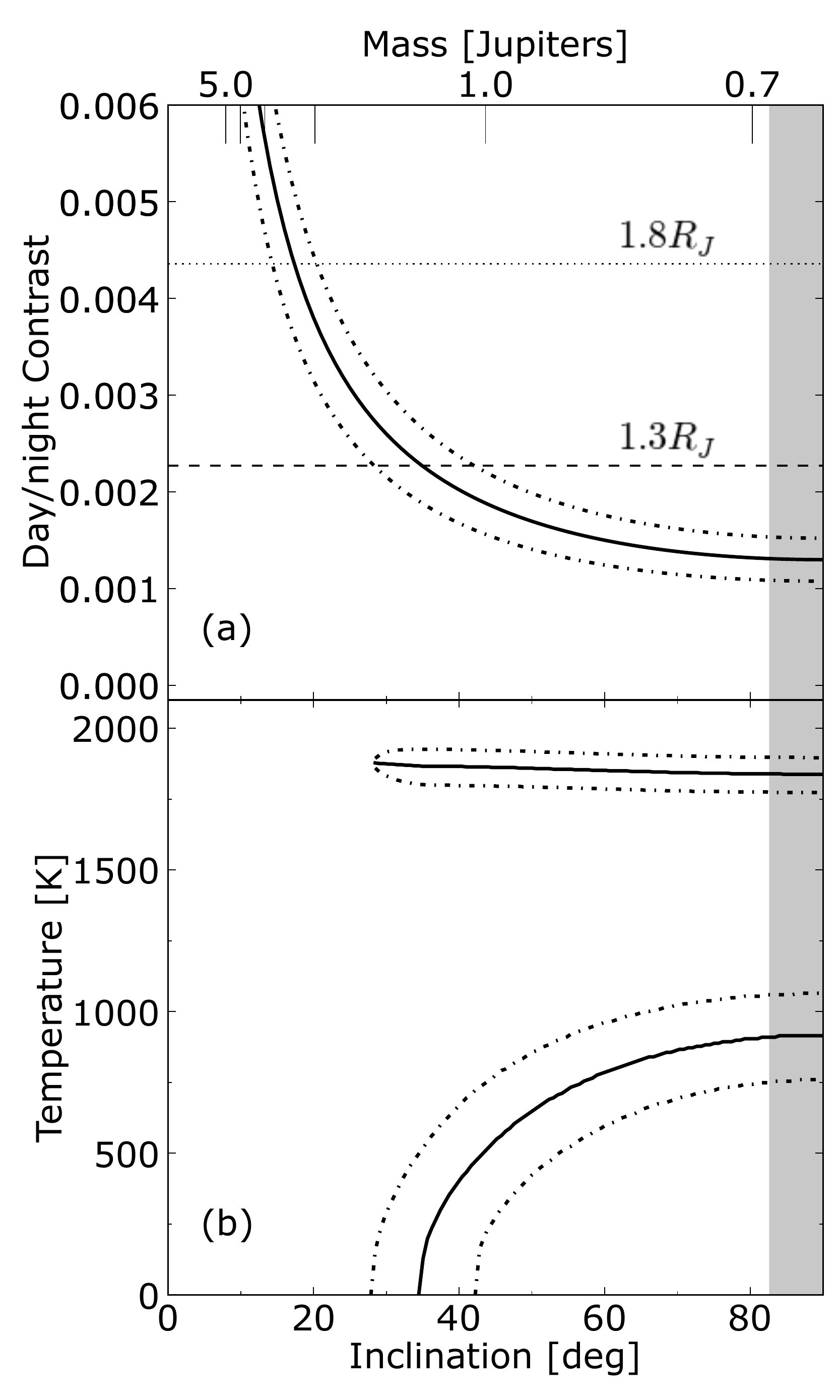}{width=4.4in}{Day-night
  contrast.}{Day-night temperature contrast as a function of orbital
  inclination angle, assuming a planet with zero albedo.  (a) Our
  measurement of the phase curve amplitude and \eref{vconstraint}
  constrain the allowed day/night contrast to lie between the
  dot-dashed 3$\sigma$ limits. The maximum allowable contrast for a
  planet of 1.3 (1.8) \rjup\ is shown as the dashed (dotted) lines,
  suggesting a lower limit on the inclination angle of $25\degg$
  ($15\degg$).  (b) Considering \eref{econstraint} and assuming a
  radius of 1.3\rjup\ allows us to determine the temperature of both
  hemispheres as a function of inclination (solid lines, with
  dot-dashed 3$\sigma$ limits).  This radius implies a planetary
  temperature contrast of $\gtrsim 900$~K.  The cooler hemisphere is
  more sensitive to changes in planetary radius, though both
  temperatures increase as radius increases.}

\fig{mr}{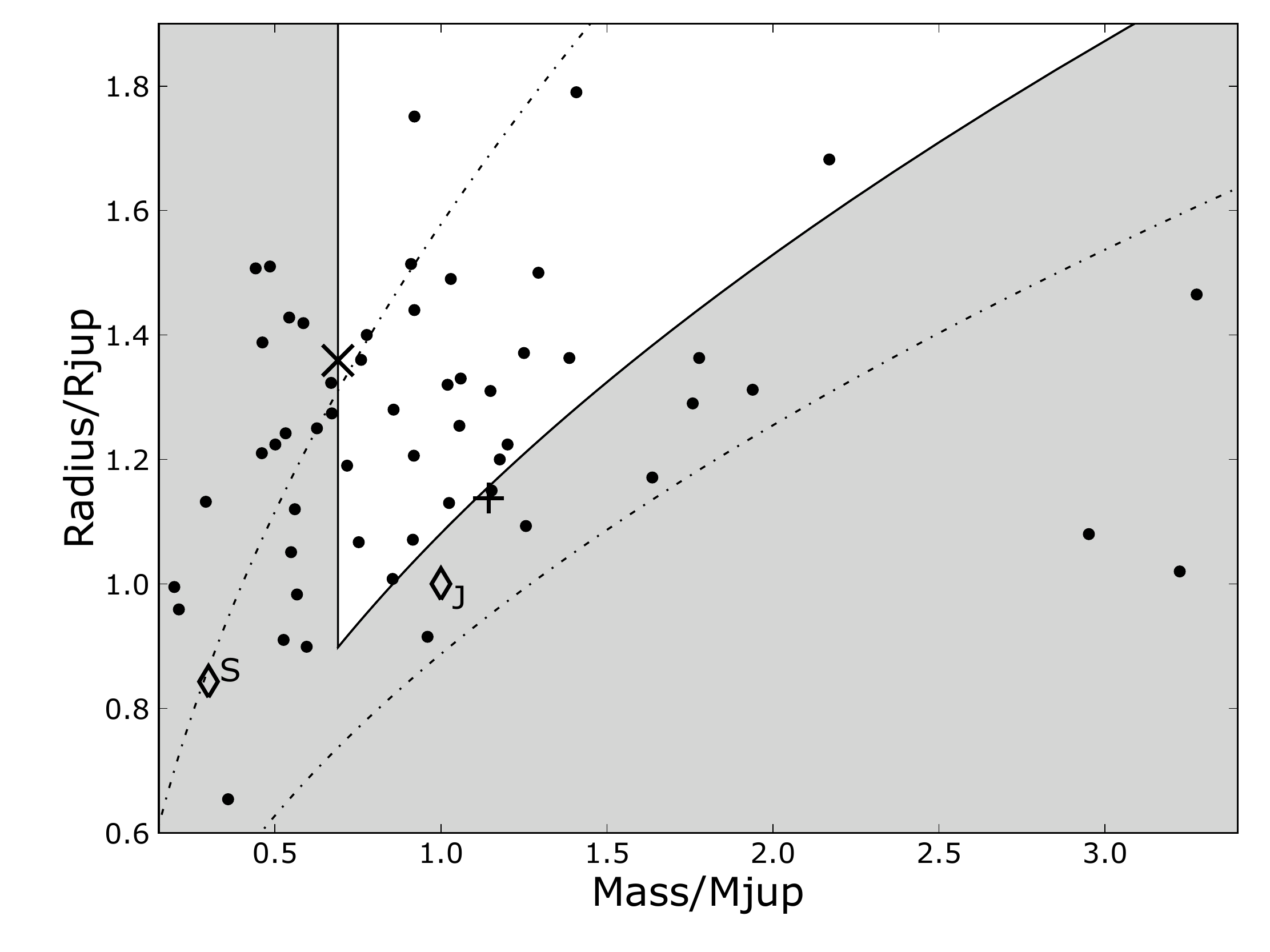}{width=5in}{\uab\ composition.}{Mass/radius
  constraints for \uab\ from \eref{maxgrav}.  The shaded area is the
  portion of mass-radius space excluded at the $3\sigma$ level.
  Points are known transiting extrasolar planets; measurement
  uncertainties have been omitted for clarity. Extrasolar planets that
  have been observed with MIPS are indicated by a `+' (\hdoneb ) and a
  `$\times$' (\hdtwob ), while Jupiter and Saturn are marked with a
  `J' and `S,' respectively. The dot-dashed lines represent lines of
  constant surface gravity with $g=10^3$~cm~s$^{-2}$ and $3.2\times
  10^3$~cm~s$^{-2}$. Our model and phase curve measurement constrain
  $g < 2100$~cm~s$^{-2}$ at the 3$\sigma$ level.}

\end{document}